\documentclass{article}
\usepackage{graphicx}
\usepackage{dcolumn}
\usepackage{bm,slashed}
\usepackage{amssymb}
\usepackage{url}
\makeatletter
\g@addto@macro{\UrlBreaks}{\UrlOrds}
\makeatother
\usepackage{float}
\usepackage{grffile}
\begin{document}
\title{Plasma-like Description for Elementary and Composite Quantum Particles}
\author{Andrey Akhmeteli\\LTASolid Inc., Houston, TX, USA\\akhmeteli@ltasolid.com}
\maketitle
\begin{abstract}
   Schr{\"o}dinger noticed in  1952 that a scalar complex wave function can be made real by a gauge transformation. The author showed recently that one real function is also enough to describe matter in the Dirac equation in an arbitrary electromagnetic or Yang--Mills field. This suggests some ``symmetry'' between positive and negative frequencies and, therefore, particles and antiparticles, so the author previously considered a description of one-particle wave functions as plasma-like collections of a large number of particles and antiparticles. The description has some similarities with Bohmian mechanics. This work offers a criterion for approximation of continuous charge density distributions by discrete ones with quantized charge based on the equality of partial Fourier sums, and an example of such approximation is computed using the homotopy continuation method. An example mathematical model of the description is proposed. The description is also extended to composite particles, such as nucleons or large molecules, regarded as collections including a composite particle and a large number of pairs of elementary particles and antiparticles. While it is not clear if this is a correct description of the reality, it can become a basis of an interesting model or useful picture of quantum mechanics.
\end{abstract}
\section{Introduction}

Recent progress in quantum information processing puts a new emphasis on foundations of quantum theory. However, it is probably safe to say that there is currently no consensus on the interpretation of quantum theory~\cite{Schloss2013,sommer,norsen,siva}. This suggests that no existing interpretation is completely satisfactory, so the formal description discussed in this work may be of some interest, if~not as a ``how actually'' model, then at least as a ``how possibly'' model (\cite{craver}, \S 3.3). The~description is intuitive; uses some notions of quantum field theory, such as vacuum polarization; and~does not seem to have problems with wave function spreading, although~it implies that the wave function has something to do with charge distribution. There are some similarities with de Broglie--Bohm interpretation (Bohmian mechanics), especially for composite particles. Bohmian mechanics is sometimes considered not just as an interpretation, but~also as another picture of quantum mechanics and a basis for computational methods~\cite{Sanz}. This can be also a way to assess the description of this~work.

There is a well-known analogy between quantum particles and plasma: the dispersion relation for the Klein--Gordon equation ($c=\hbar=1$)
\begin{equation}\label{eq:kgdr}
\omega^2=m^2+k^2
\end{equation}
is similar to the dispersion relation for waves in a simple plasma model
\begin{equation}\label{eq:pdr}
\omega^2=\omega_p^2+k^2.
\end{equation}
However, to~expand this analogy, we need a description of both negatively and positively charged particles.The description is based on the little-known possibility of using a real, rather than complex, wave function to describe charged particles. Schr{\"o}dinger~\cite{Schroed} noticed that a charged scalar field can be made real by a gauge transformation, in spite of ``the widespread belief about `charged’ fields requiring complex representation.'' As a generalization of this result, it was shown recently (see in \cite{Akhmeteli-JMP,Akhm2015,Akhmspr}; see also~in \cite{Bagrov2014}, pp. 24--25,~\cite{Bagro}) that, in~a general case, one can use just one real function instead of the four complex components of the Dirac spinor in the Dirac equation in an arbitrary electromagnetic field at the expense of getting a fourth-order partial differential equation. A~similar result can be derived for the Dirac equation in a non-Abelian gauge field~\cite{Akhmeteliym}.

Using one real wave function instead of complex functions introduces some ``symmetry'' between positive and negative frequencies and, thus, particles and antiparticles. Therefore, a~tentative description of such (one-particle electron) wave function  was given in~\cite{Akhm10,Akhmeteli-EPJC}: the wave function can describe $N+1$ electrons and $N$ positrons, where $N$ is very large. This collection of particles and antiparticles has the same total charge (and mass) as an electron, and~the value of the wave function at some point (or some combination of the wave function and its derivatives at the point) is a measure of both  ``vacuum polarization'' and the density of probability of finding an electron at this point (finding a positron at that point is also possible, but~probably requires much more energy). An~electron found during a measurement can be any of the $N+1$ electrons. The~results of the measurement on the specific collection can depend, say, on~the precise coordinates of the particles in the collection. One can consider such a collection as a ``dressed'' electron with a well-defined total charge. The~description assumes trajectories of the ``bare'' electrons and positrons, but~the uncertainty principle is valid for the ``dressed'' electron. The~charge density distribution of the ``dressed'' electron is defined by all charges of the ``bare'' electrons and positrons and can be very close to the charge density distribution built from the traditional wave function (see Section~\ref{s21}). If~an electron is removed from the collection (for example, a~precise position measurement means high momentum uncertainty, as~a result, some particle acquires high speed, quickly leaves the collection, and~the area around the place vacated by the particle will tend to be filled with the surrounding particles) and the energy of the remaining collection is not high enough for the collection to manifest as pairs, it is difficult to tell the collection from vacuum. This may be the source of discreteness emphasized in~\cite{Khentropy}. It is also important to note that spreading of wave packets (which complicates, e.g.,~the de Broglie's double solution approach~\cite{colin7}) is not problematic for this~description.

This description suggests an analogy between plasma and elementary particles. Such analogy was discussed long ago. For~example, Vigier~\cite{Vig} considered a vacuum containing fermionic and antifermionic fields, and compared it to a plasma (see also references in Section \ref{s3}).

It is difficult to say if the vast body of work on quark--gluon plasma is directly relevant to interpretation of quantum theory as quark-gluon plasma is a high-temperature or high-density state of matter~\cite{Braun}.

The description of this article only covers the unitary evolution of quantum theory, but~not the wave function collapse. On~the one hand, the~author would like to essentially limit the discussion to matters of mathematical physics, on~the other hand, there are now some indications that unitary evolution may be sufficient to describe experiments. For~example, Schlosshauer reviewed experimental data and concluded~\cite{Schloss}: ``(i) the universal validity of unitary dynamics and the superposition principle has been confirmed far into the mesoscopic and macroscopic realm in all experiments conducted thus far; (ii) all observed ``restrictions'' can be correctly and completely accounted for by taking into account environmental decoherence effects; (iii) no positive experimental evidence exists for physical state-vector collapse''. Furthermore, Allahverdyan e.a. theoretically studied dynamics of a spin interacting with a quantum model of a measuring apparatus and concluded~\cite{ABN} that ``uniqueness of the outcome of each run and reduction can be derived from the Hamiltonian dynamics of the macroscopic pointer alone'', and~``recurrences might still occur, but~at inaccessibly large~times''.

One can object that the mass of such a collection of a large number of particles and antiparticles would be too large, as~each pair should have a mass of at least two electron masses, but~it is not necessarily so, as~the energy of an electron and a positron that are very close to each other can be significantly less than their energy when the distance between them is~large.

\begin{figure}[ht]
 \begin{tabular}{cc}
\includegraphics[width=0.45\linewidth]{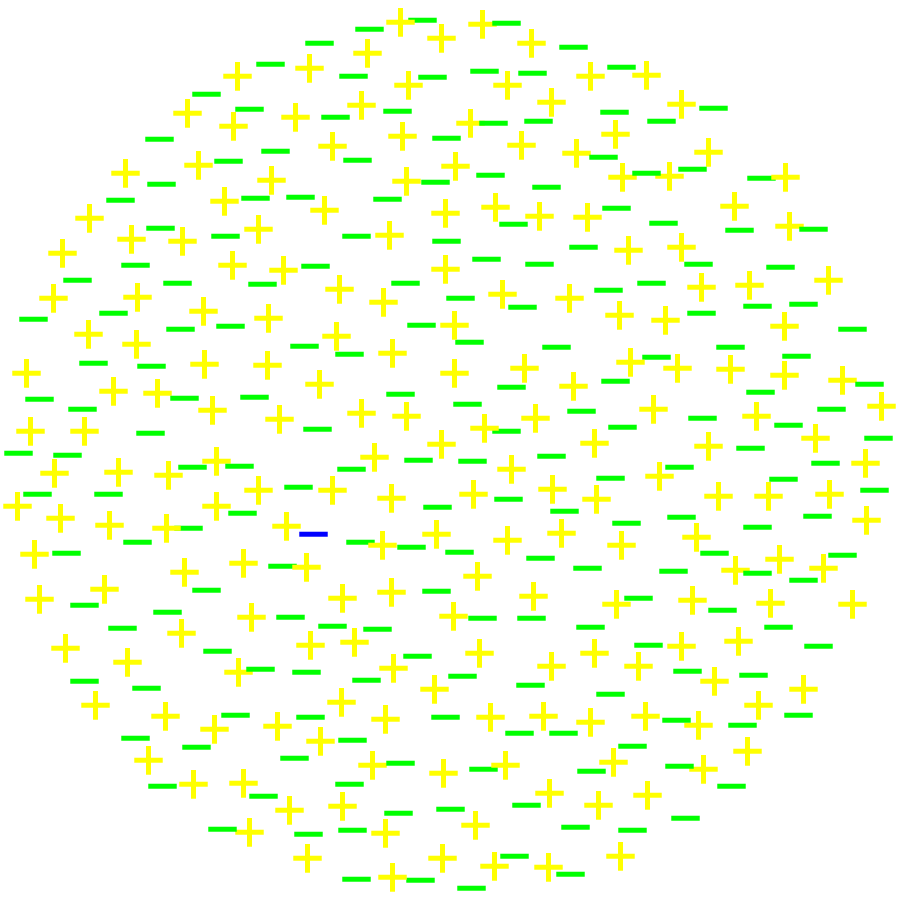}
&\includegraphics[width=0.45\linewidth]{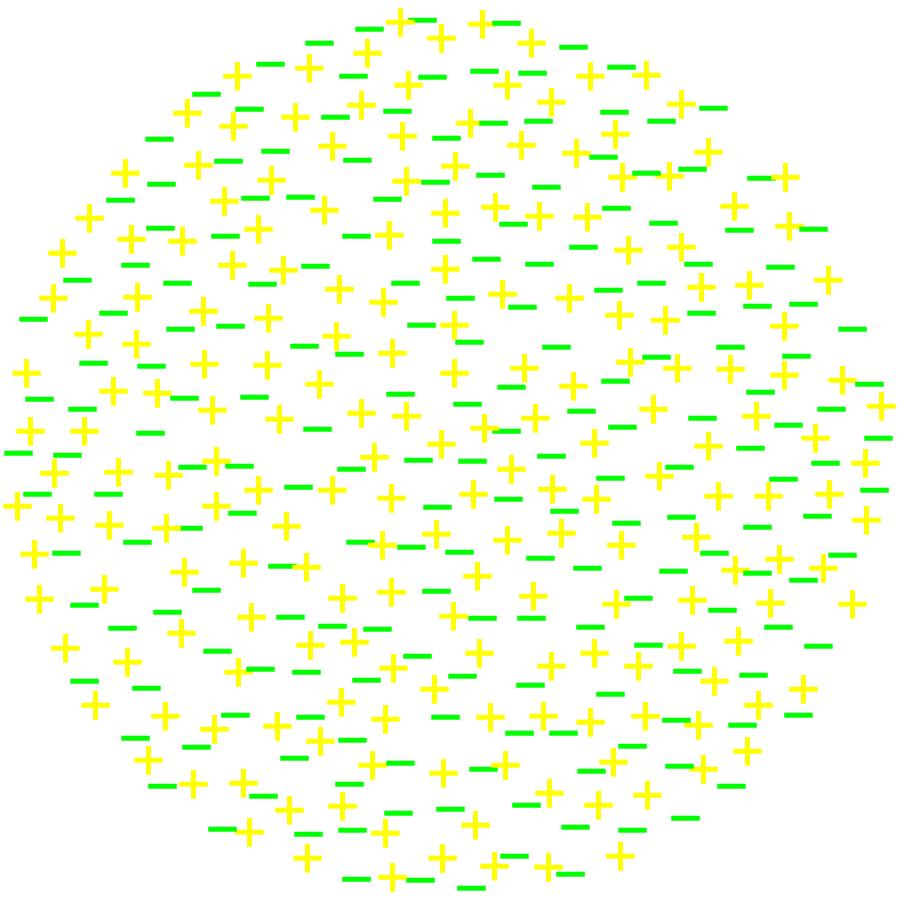}\\
({\bf a})&({\bf b})\\
\end{tabular}
\caption{Collections (\textbf{a},\textbf{b}) have 201 and 200 electrons, respectively, and~200 positrons~each.}
\label{fig:pm}
\end{figure}

The above description is illustrated by Figure~\ref{fig:pm}, where electrons and positrons are represented by minus and plus signs, respectively. Collections (a) and (b) there are identical except for an extra electron (represented by a blue minus sign) in collection (a). The~collections are difficult to visually tell from each other, but~the total charges of the collections are one and zero electron charge, respectively. Figure~\ref{fig:pm} is similar to M. Strassler's Figure~3 
 at~\cite{Strassler}, but~the figure here describes an electron, rather than a nucleon (which is a composite particle), and~the size of the collection is defined by the size of the volume where the wave function does not~vanish.

The description has obvious similarities with the Bohmian interpretation. Let us discuss them for a single-particle system (as it is understood both traditionally and in the Bohmian interpretation~\cite{Bohmhil}, because~there are always numerous particles in the plasma-like description). If~the system is described by a wave function, the~latter defines current lines, which can be regarded as possible particle trajectories in both the Bohmian interpretation and the plasma-like description. There is only one particle and one trajectory in the Bohmian interpretation (other trajectories are realized in other instances of the statistical ensemble~\cite{Bohmhil}), and there are numerous particles and trajectories in the plasma-like description. However, there is more similarity in the case of a composite particle, where the trajectory of the composite particle itself seems singled-out/different from the trajectories of the surrounding elementary fermions/antifermions in the plasma-like description (see Section~\ref{s3}).

While single-particle systems are very important (for example, they are sufficient to describe the double-slit experiment), it is necessary to discuss many-particle systems (cf. V. Vedral's comments~\cite{Vedral} on the approach in~\cite{Akhmeteli-IJQI}). The~author does not have a complete description of second quantization for such systems, but~for fermions the Pauli principle can emerge in the following way: for identical wave functions, the~relevant collections of discrete charges have identical or very similar coordinates of the charges and thus a combination of such collections can have a very high energy, for~example, due to Coulomb interaction. Note that matter and radiation (and, eventually, Fermi and Bose fields) are treated differently in the plasma-like description. Second quantization for bosons can be performed using the same approach as in~\cite{Akhmeteli-EPJC}, Section~4 
 (the approach using a generalized Carleman linearization~\cite{Kowalski,Kowalski2} was proposed by nightlight): for a rather general system of nonlinear differential equations of evolution in a 3-dimensional space, the~set of solutions can be embedded into the set of solutions of a system of linear differential equations of evolution in the Fock space with a Hamiltonian built using boson~operators.

In this article, we offer a tentative resolution of some important issues arising in this description. In~particular, we try to answer the following~questions:
\begin{itemize}
	\item How can a continuous charge density distribution be approximated  by a discrete one with a quantized charge?
    \item What mathematical model (equations of motion) can underlie such a description?
    \item How can the description be extended to composite particles?
	\item What are some implications of the obvious analogy with plasma?
\end{itemize}

\section{Methods and~Results}\label{s2}

\subsection{Approximation of a Continuous Charge Density Distribution by a Discrete One with a Quantized~Charge}\label{s21}

How accurately can a continuous charge density distribution for a specific wave function with a total charge equal to one electron charge be approximated by a collection of discrete charges with values of $\pm 1$ electron charge? It is obvious that a Fourier expansion of a point-like charge density distribution contains arbitrarily high spatial frequencies, whereas high-spatial-frequency Fourier components of smooth charge density distributions tend to zero fast; therefore, it is probably impossible to approximate a continuous charge density distribution by a finite number of discrete quantized charges with a good accuracy. However, quantum field theories are typically considered to be just effective theories: ``...we are now used to the idea that there are important interactions at many different energy scales, some of them probably so large that we cannot see them, directly. Certainly not now. Perhaps not ever. Nevertheless, we can use an effective field theory to describe physics at a given energy scale, $E$, to~a given accuracy, $\epsilon$, in~terms of a quantum field theory with a finite set of parameters. We can formulate the effective field theory without any reference to what goes on at arbitrarily small distances~\cite{georgi}.'' Therefore, we can look for collections of discrete charges with values of $\pm 1$ electron charge that have the same Fourier components with spatial frequencies below some limit value as the initial smooth charge density~distribution.

\begin{figure}[ht]
\includegraphics[scale=0.5,width=.4\linewidth]{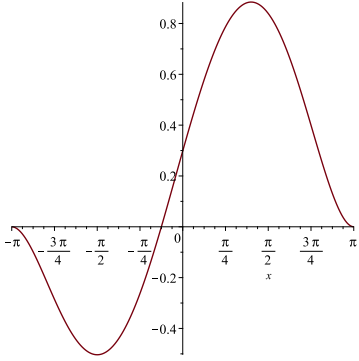}
\caption{The smooth charge density distribution to be approximated by a collection of discrete quantized~charges.} 
\label{fig:fx2}

\end{figure}

Let us illustrate this approach in the one-dimensional case. We assume that the smooth charge density distribution is periodic, e.g.,~with a period of $2\pi$, so we consider it on a segment $[-\pi,\pi]$. Let us choose the following function for our example:
\begin{equation}\label{eq:fx}
f(x)=\frac{15}{2\pi^6}(x^2-\pi^2)^2(x+\frac{\pi}{8})
\end{equation}
(see Figure~\ref{fig:fx2}). Let us note that
\begin{equation}\label{eq:norm}
\int_{-\pi}^\pi f(x)dx=1
\end{equation}
(the total charge in the charge density distribution is +1 (electron charge)).
The charge density distribution was chosen not to be non-negative everywhere, as~we have in mind applications not just to the Schr{\"o}dinger equation and the Dirac equation, but~also to the Klein--Gordon equation, where the charge density does not have to be of the same sign~everywhere.

Let us consider the Fourier expansion:
\begin{equation}\label{eq:four}
f(x)=\frac{1}{2}a_0+\sum_{k=1}^\infty(a_k \cos (k x)+b_k \sin (k x)),
\end{equation}
where
\begin{equation}\label{eq:fourser}
a_k=\frac{1}{\pi}\int_{-\pi}^\pi f(y)\cos (k y) dy,\;\;b_k=\frac{1}{\pi}\int_{-\pi}^\pi f(y)\sin (k y) dy.
\end{equation}

We will try to find a discrete charge density distribution approximating $f(x)$. Let us assume that this distribution describes $2j+1$ particles with coordinates $x_n$, including $j+1$ electrons and $j$ positrons, so the discrete charge density distribution is
\begin{equation}\label{eq:gx}
g(x)=\sum_{n=1}^{2j+1}(-1)^{n+1}\delta(x-x_n).
\end{equation}

Let us demand that the $k$-th cosine Fourier components of distribution $g(x)$ coincide with $a_k$ for $1\leq k\leq k_c$ (the zeroth cosine Fourier component of $g(x)$ automatically coincides with $a_0$ due to Equation~(\ref{eq:norm})) and that the $k$-th sine Fourier components of distribution $g(x)$ coincide with $b_k$ for $1\leq k\leq k_s$, where $k_c$ and $k_s$ are some natural~numbers.

Let us introduce the following notation: $u_n=\cos(x_n), v_n=\sin(x_n)$. As~\begin{eqnarray}\label{eq:cs}
\nonumber
\cos(k x_n)=\frac{1}{2}((\cos(x_n)+i\sin(x_n))^k+(\cos(x_n)-i\sin(x_n))^k)=
\\
\nonumber
\frac{1}{2}((u_n+iv_n)^k+(u_n-iv_n)^k),
\\
\nonumber
\sin(k x_n)=\frac{1}{2i}((\cos(x_n)+i\sin(x_n))^k-(\cos(x_n)-i\sin(x_n))^k)=
\\
\frac{1}{2i}((u_n+iv_n)^k-(u_n-iv_n)^k),
\end{eqnarray}
$\cos(k x_n)$ and $\sin(k x_n)$ can be expressed as polynomials of $u_n$ and $v_n$, and~the equality of $k_c+k_s$ Fourier components of $g(x)$ to Fourier components $a_k$ and $b_k$ of $f(x)$ can be expressed  as $k_c+k_s$ polynomial equations for $u_n$ and $v_n$. Adding equations $u_n^2+v_n^2=1$, we obtain a system of $k_c+k_s+2j+1$ polynomial equations for $2(2j+1)$ unknowns $u_n$ and $v_n$:
\begin{eqnarray}\label{eq:ps}
\nonumber
\frac{1}{\pi}\sum_{n=1}^{2j+1}(-1)^{n+1}\frac{1}{2}((u_n+iv_n)^k+(u_n-iv_n)^k)=a_k,\;\; 1\leq k\leq k_c,
\\
\nonumber
\frac{1}{\pi}\sum_{n=1}^{2j+1}(-1)^{n+1}\frac{1}{2i}((u_n+iv_n)^k-(u_n-iv_n)^k)=b_k,\;\; 1\leq k\leq k_s,
\\
u_n^2+v_n^2=1,\;\; 1\leq n\leq 2j+1.
\end{eqnarray}

It seems likely (but the author has not proven) that, given some values of $k_c$ and $k_s$, this system will always have some real solutions for sufficiently large $j$, or, in~other words, it is always possible to find a discrete charge density distribution with a sufficiently large number of point-like particles that has the same $k_c+k_s$ Fourier components as the initial smooth charge density distribution. However, the~polynomial system is very complicated for large values of $k_c,k_s,$ and $j$, and~it is difficult to find such discrete distribution. It is possible though to do that numerically, using the powerful homotopy continuation method for polynomial systems (see~\cite{vers} and references there). The~idea of homotopic continuation is as follows \cite{leykin}
. If~there is a square system of polynomial equations $F=(f_1(x),...,f_m(x))$, where $x=(x_1,...,x_m)$, $f_1(x),...,f_m(x)$ are the left-hand sides of the equations, and~the right-hand sides equal zero, one uses the homotopy
\begin{equation}\label{eq:ht}
H_t=(1-t)G+tF,\;\; t\in[0,1],
\end{equation}
connecting a start system $G=H_0$ with the target system $F=H_1$. If~we differentiate $H_t(x(t))=0$ with respect to $t$, we obtain
\begin{equation}\label{eq:ode}
\left( \frac{\partial H_t}{\partial x}x'(t)+\frac{\partial H_t}{\partial t} \right)_{x=x(t)}=0.
\end{equation}
This yields a system of ordinary differential equations for $x(t)$, which can be solved numerically, if~one knows a solution of the start~system.

While the typical goal of the homotopy continuation method is finding all the solutions of the target polynomial system, our goal is quite limited: we would like to find just one solution of the target system to get an idea of how a discrete charge density distribution with a quantized charge can approximate a smooth distribution. For~this reason, we build the start system $G$ and its solution as follows. We generate $2j+1$ random values $\tilde{x}_n$, such that $-\pi\leq\tilde{x}_n\leq\pi$, and~obtain a sequence $\bar{x}_n$ by sorting $\tilde{x}_n$ in ascending order. Then, we construct sequences $\bar{u}_n=\cos(\bar{x}_n)$ and $\bar{v}_n=\sin(\bar{x}_n)$ and obtain the start system by replacing $a_k$ and $b_k$ in the right-hand sides of the first $k_c+k_s$ equations of system (\ref{eq:ps}) by the values of their left-hand sides after substitution $u_n=\bar{u}_n,v_n=\bar{v}_n$. Obviously, $\bar{u}_n,\bar{v}_n$ is a solution of the start system, so we can build a solution of system (\ref{eq:ps}) using homotopy~continuation.

We obtained such a solution for $k_c=25,k_s=24,j=24$ (the target and start systems are square for these values, with~98 unknowns and 98 equations, each) using the solver from ~\cite{vers} (the latest version and documentation can be found at~\cite{vers2}). The~results are presented in Figure~\ref{fig:res}. The~calculated coordinates of the discrete charges are given in the Appendix. Note that the partial Fourier sums coincide for the smooth and discrete charge density distributions. Let us also note that the difference between the smooth distribution and its partial Fourier sum is less than $0.1\%$ of the maximum value of the distribution. Empirically, the~charges are mostly arranged in pairs; clusters of three charges are denoted by the ellipses (this choice of the clusters is rather arbitrary, but~may be useful as an illustration). It is interesting that the solution of the start system was generated by choosing random coordinates between $-\pi$ and $\pi$ and did not display many pairs/clusters. The~pairs/clusters in the solution of the target system probably appeared because the charge density distribution to be approximated was~smooth.

The approach of this section can be useful for interpretation of quantum phase-space distribution functions~\cite{lee}, such as the Wigner distribution function, which are not necessarily non-negative. It is also interesting to compare the approach with the initial Schrödinger's interpretation of the wave function (see, e.g.,~~in \cite{Sebens,Barutschr} and references therein). Schrödinger's interpretation of $e|\psi|^2$ as charge density meets some objections. For example, A. Khrennikov noted (\cite{Khbq}, p.~23) :``Unfortunately, I was not able to find in Schrödinger's papers any explanation of the impossibility to divide this cloud into a few smaller clouds, i.e.,~no attempt to explain the fundamental discreteness of the electric charge.'' The plasma-like description suggests that $e|\psi|^2$ (and its analogs for the Klein--Gordon and Dirac equations) is just a smoothed charge density, and~the description is immune to the above~objection.
\begin{figure}[ht]
\includegraphics[scale=1,width=1\linewidth]{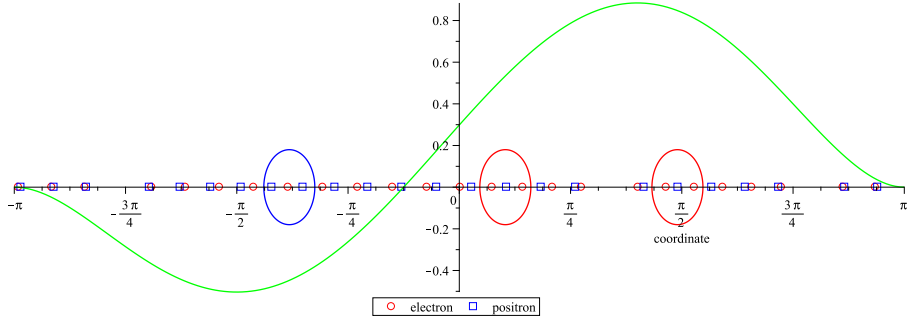}
\caption{The smooth charge density distribution (green) and the discrete charges of the approximating discrete charge density distribution. The~charges are mostly arranged in pairs; the clusters of three charges are encircled by~ellipses. 
 }
\label{fig:res}

\end{figure}
\unskip
\subsection{An Example Mathematical~Model}\label{s22}

Let us define an example mathematical model (equations of motion) of collections of charged particles and antiparticles interacting with electromagnetic field. As~we want the experimental predictions of such a model to be as close as possible to those of quantum theory, we will use scalar electrodynamics (the Klein--Gordon--Maxwell electrodynamics) as our starting point. Its equations of motion are as follows~\cite{Schroed}:
\begin{equation}\label{eq:pr7}
(\partial^\mu+ieA^\mu)(\partial_\mu+ieA_\mu)\psi+m^2\psi=0,
\end{equation}
\begin{equation}\label{eq:pr8}
\Box A_\mu-A^\nu_{,\nu\mu}=j_\mu,
\end{equation}
\begin{equation}\label{eq:pr9}
j_\mu=ie(\psi^*\psi_{,\mu}-\psi^*_{,\mu}\psi)-2e^2 A_\mu\psi^*\psi.
\end{equation}
The metric signature is $(+,-,-,-)$, and~$\Box=\partial^\mu\partial_\mu$ is the~d'Alembertian.

The complex charged matter field $\psi$ in scalar electrodynamics (Equations (\ref{eq:pr7})--(\ref{eq:pr9})) can be made real by a gauge transformation (at least locally), and~the equations of motion in the relevant gauge (unitary gauge) for the transformed four-potential of electromagnetic field $B^{\mu}$ and real matter field $\varphi$ are as follows~\cite{Schroed}:
\begin{equation}\label{eq:pr10}
\Box\varphi-(e^2 B^\mu B_\mu-m^2)\varphi=0,
\end{equation}
\begin{equation}\label{eq:pr11}
\Box B_\mu-B^\nu_{,\nu\mu}=j_\mu,
\end{equation}
\begin{equation}\label{eq:pr12}
j_\mu=-2e^2 B_\mu\varphi^2.
\end{equation}

Using a substitution $\Phi=\varphi^2$, one can obtain~\cite{Akhmeteli-EPJC}
\begin{equation}\label{eq:pr3q}
\Box\Phi-\frac{1}{2}\frac{\Phi^{,\mu}\Phi_{,\mu}}{\Phi}-2(e^2 B^\mu B_\mu-m^2)\Phi=0,
\end{equation}
\begin{equation}\label{eq:pr11q}
\Box B_\mu-B^\nu_{,\nu\mu}=j_\mu,
\end{equation}
\begin{equation}\label{eq:pr12q}
j_\mu=-2e^2 B_\mu\Phi.
\end{equation}

Let us start building the model emulating scalar electrodynamics. We assume that we have $N+1$ particles with charge $e$ and $N$ antiparticles with charge $-e$. The~trajectory of the n-th particle/antiparticle is $\boldsymbol{x}_n(x^0)$. Thus, the~charge density distribution is
\begin{equation}\label{eq:g1}
g^0(x^0,\boldsymbol{x})=e\sum_{n=1}^{2N+1}(-1)^{n+1}\delta(\boldsymbol{x}-\boldsymbol{x}_n(x^0)).
\end{equation}
Here, $\boldsymbol{x}_n(x^0)=(x^1_n(x^0),x^2_n(x^0),x^3_n(x^0))$ is the 3D coordinate of the n-th particle/antiparticle at the time point $x^0$. Greek indices run from 0 to 3 and Latin indices run from 1 to 3, unless~they denote the particle number. Particles have odd numbers, and~antiparticles have even numbers.
Let us present $g^0(x)$ as a 3D Fourier integral:
\begin{equation}\label{eq:g3}
g^0(x^0,\boldsymbol{x})=(2\pi)^{-\frac{3}{2}}\int G^0(x^0,\boldsymbol{k})\exp(i \boldsymbol{k} \boldsymbol{x})d \boldsymbol{k},
\end{equation}
where
\begin{equation}\label{eq:g4}
G^0(x^0,\boldsymbol{k})=(2\pi)^{-\frac{3}{2}}\int g^0(x^0,\boldsymbol{x})\exp(-i \boldsymbol{k} \boldsymbol{x})d \boldsymbol{x}.
\end{equation}
Let us now define a smoothed initial charge density at $x^0=x^0_{in}$:
\begin{equation}\label{eq:g5}
g^0_{sm}(x^0_{in},\boldsymbol{x})=(2\pi)^{-\frac{3}{2}}\int_{|\boldsymbol{k}|<\lambda} G^0(x^0_{in},\boldsymbol{k})\exp(i \boldsymbol{k} \boldsymbol{x})d \boldsymbol{k},
\end{equation}
where $\lambda>0$ is a large cutoff~constant.

Let us construct the smoothed initial current density at $x^0=x^0_{in}$ as follows (the smoothing process is chosen rather arbitrarily, but~based on the requirement that the directions of the 4-velocities of the particles and the 4-potentials of the electromagnetic field in the same point coincide; up to a factor, the~smoothed charge density is a sum of smoothed delta-functions describing particles):
\begin{equation}\label{eq:g6}
g^i_{sm}(x^0_{in},\boldsymbol{x})=\frac{B^i(x^0_{in},\boldsymbol{x})}{B^0(x^0_{in},\boldsymbol{x})}g^0_{sm}(x^0_{in},\boldsymbol{x}).
\end{equation}
Now, let the initial 4-current density be equal to the smoothed 4-current density:
\begin{equation}\label{eq:g7}
j^\mu(x^0_{in},\boldsymbol{x})=g^\mu_{sm}(x^0_{in},\boldsymbol{x}).
\end{equation}
Let us assume that $B^\mu(x^0,\boldsymbol{x})$ and the temporal derivatives $\dot{B^\mu}(x^0,\boldsymbol{x})$ are defined everywhere in the 3D space for $x^0=x^0_{in}$ in such a way that
\begin{equation}\label{eq:g8}
\Box B_0(x^0_{in},\boldsymbol{x})-B^\nu_{,\nu 0}(x^0_{in},\boldsymbol{x})=B^{,i}_{0,i}(x^0_{in},\boldsymbol{x})-\dot{B}^i_{,i}(x^0_{in},\boldsymbol{x})=j^0(x^0_{in},\boldsymbol{x}).
\end{equation}
As was shown in~\cite{Akhmeteli-EPJC}, Section~2, 
 if~$B^{\mu}$ and $\dot{B}^{\mu}$ are defined in the entire 3D space at some point in time $x^0$, $\ddot{B}^{\mu}(x^0,\boldsymbol{x})$ can be calculated from Equations~(\ref{eq:pr3q})--(\ref{eq:pr12q}), so the Cauchy problem can be posed and $B^{\mu}$ can be calculated in the entire spacetime (we use the following notation: $\dot{Y}=Y^{,0}$ and $\ddot{Y}=Y^{,00}$ are the first and second temporal derivatives of Y, correspondingly). The~trajectories of the particles/antiparticles can be calculated using the condition $\boldsymbol{\dot{x_n}}(x^0)=\frac{\boldsymbol{B}(x^0,\boldsymbol{x_n})}{B^0(x^0,\boldsymbol{x_n})}$.

On the one hand, the~above equations of motions coincide with those of scalar electrodynamics for some choice of initial conditions, on~the other hand, one can expect that for a large cut-off constant the current will be close to that of a collection of point~particles.

The current in this example model is not always time-like, but~this problem is inherited from scalar electrodynamics and can probably be overcome if we start with spinor electrodynamics (Dirac--Maxwell electrodynamics). This issue was discussed for the de Broglie--Bohm interpretation of the Klein--Gordon field \cite{Nik1,Dewd1,Tum2,Holland}. In~the present article, however, the~Klein--Gordon equation is regarded just as a reasonably decent approximation for~electrons.

As the model of this section emulates arbitrarily well (for a sufficiently large cut-off constant) a quantum theory---the Klein--Gordon electrodynamics (scalar electrodynamics), one can reasonably expect that the model should successfully describe a wide spectrum of quantum phenomena. Modeling of specific experiments, such as the double-slit experiment, is left for future~work.

It is not clear if this model can describe pair creation or annihilation. One probably needs to study continuation of the solutions at the points $x$ where $B^0(x)=0$.

Let us emphasize again that one needs to choose the cutoff constant to fully define the mathematical model. While this constant can be arbitrarily high, the~model has problems at very high temporal/spatial frequencies once this choice is made. However, these problems seem similar to those of standard quantum field theories~\cite{georgi}.

In the example mathematical model, the~electromagnetic field guides numerous particles/antiparticles, whereas the particles/antiparticles act as a source of the electromagnetic field. In~comparison, previously the author showed that the Klein--Gordon--Maxwell electrodynamics in the unitary gauge allows natural elimination of the particle wave function and describes independent evolution of the electromagnetic field. Therefore, the~electromagnetic field can be regarded as the guiding field in the Bohmian interpretation~\cite{Akhm10,Akhmeteli-IJQI,Akhmeteli-EPJC}.

\subsection{Extension to Composite~Particles}\label{s3}
\begin{figure}[ht]
\includegraphics[scale=0.6,width=.5\linewidth]{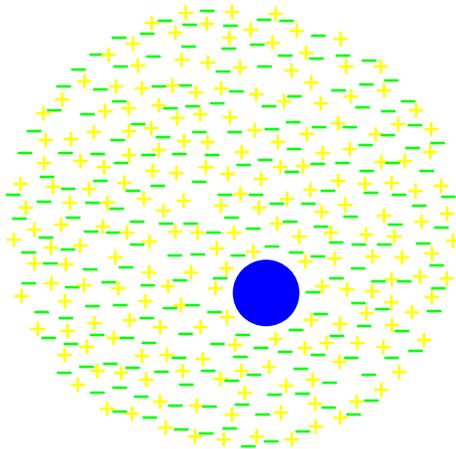}
\caption{A composite particle (the blue disk) and a collection of~pairs.}
\label{fig:comp}
\end{figure}

Let us try to resolve the following problem of the description. Composite particles, such as nucleons or large molecules, also demonstrate quantum properties~\cite{zeilneut,Fein}. It is however difficult to imagine that molecule--antimolecule pairs play a significant role in diffraction of large molecules (creation of such pairs is possible, but~much less probable than creation of electron--positron pairs). However, composite particles take part in some interactions (for example, electromagnetic or strong interactions), so the description can be modified as follows in that case: composite particles are accompanied by a large collection of fermion--antifermion pairs (for example, electron-positron pairs for electromagnetic interactions and quark--antiquark pairs for strong interactions; in some situations, it can be difficult to tell such pairs from force carriers, such as photons or gluons). Such fermions/pairs/force carriers are present at all locations where the wave function traditionally describing the composite particle does not vanish, so the dimensions of the collection are not limited by the range of the interaction (for example, the~short range of strong interaction). Thus, the~composite particle can be detected at all locations where the wave function does not vanish, although~at most locations it is fermions/pairs/force carriers of the collection that interact directly with the instrument, not the composite particle itself. Such a composite particle with a collection of pairs is illustrated by Figure~\ref{fig:comp}.

This does not mean that composite particles require an approach that is fundamentally different from that for elementary particles. Fundamentally, composite particles consist of elementary particles, which can be described as collections of particles and antiparticles, but~some part of them forms a bound state, so the ``bare'' composite particle (the blue disk in Figure~\ref{fig:comp}) retains its individuality. Thus, Figure~\ref{fig:comp} is just a higher-level (less detailed) picture.

Let us note that in the processes of particle diffraction and interference, the~momentum transfer is defined by Fourier components of matter distribution in the crystal lattice/diffraction grating/screen~\cite{Duane,Ehren,Lande}, so the mass of the incident particle plays a ``passive'' role: for the same momentum transfer, the~effective de Broglie wavelength is shorter for a particle of a larger~mass.

Such description is closer to the de Broglie--Bohm interpretation and the Couder experiment~\cite{Couder} for composite particles than, e.g.,~for electrons, as~a composite particle is different from other particles in the collection, and~it seems natural to single out its trajectory from all trajectories of the particles in the~collection.

\maketitle

\subsection{The Plasma~Analogy}\label{s24}
There is an obvious analogy between this description and plasma. As~the dispersion relation for the Klein--Gordon equation $\omega^2=m^2+k^2$ (in a natural system of units) is similar to a dispersion relation of a simple plasma model, such analogy was used previously (see, e.g.,~in \cite{Vig,Plyu,Shi}). This analogy illustrates the effective long-range interaction within a~collection.

Let us try to use this analogy to get an idea of the density of particles in a collection modeling what is perceived as one particle in traditional quantum experiments. If~$n_e$ is the electron density in the collection, the~plasma frequency $\omega_p$ in the electron--positron plasma is $\sqrt{2}$ times greater than the traditional plasma frequency~\cite{stenson}, i.e.,
\begin{equation}\label{eq:pf}
\omega_p=\sqrt{\frac{8\pi n_e e^2}{m_e}}
\end{equation}
(we do not consider any renormalization of mass and charge in this preliminary treatment). It is natural to suggest that this plasma frequency is equal (maybe on the order of magnitude) to the angular frequency of Zitterbewegung $\frac{2m_e c^2}{\hbar}$ ~\cite{Diraczit,thallerzit}, so we obtain
\begin{equation}\label{eq:pf2}
n_e=\frac{m_e^3 c^4}{2\pi\hbar^2 e^2}=\left(\frac{m_e c}{\hbar}\right)^3\frac{c\hbar}{2\pi e^2}=\frac{1}{2\pi\alpha}\left(\frac{\hbar}{m_e c}\right)^{-3},
\end{equation}
where $\alpha=\frac{e^2}{\hbar c}$ is the fine structure constant. Thus, $n_e\approx 3.8\cdot10^{32} cm^{-3}$ or 21.8 per cube with an edge length equal to the reduced Compton wavelength $\frac{\hbar}{m_e c}\approx 3.86\cdot 10^{-11}\textrm{cm}$. The~high electron density suggests that there is low energy per particle of a collection. Let us also note that in this context the Zitterbewegung frequency plays a role of a ``natural frequency'', rather than a frequency of some ``internal clock''~\cite{Gou}.

The plasma analogy suggests that more complex equations of motion can be useful for quantum theory, e.g.,~some analogs of the Vlasov~equation.


\section{Conclusions}\label{s4}

We considered a possible description of one-particle wave functions as plasma-like collections of particles and antiparticles, and proposed an approach to approximating smooth charge density distributions by discrete ones with quantized charge based on the requirement that partial Fourier sums are equal for the initial and the approximating distributions. An~example of an approximating discrete distribution was computed using the homotopy continuation method for polynomial~systems.

An example mathematical model based on the Klein--Gordon--Maxwell electrodynamics (scalar electrodynamics) is~proposed.

The description was extended to composite particles, and~some implications of the plasma analogy were~derived.

One cannot be sure that this description correctly describes reality, but~even if it does not, it provides an interesting model or useful picture of quantum mechanics and an approach to understanding quantum~randomness.

\section*{Funding}
This research received no external~funding.
\section*{Acknowledgements}
The author is grateful to A. V. Gavrilin, A. Yu. Kamenshchik, T. G. Khunjua, A. D. Shatkus, and~ A. Tarasevitch for their interest in this work and valuable remarks. The~author is also grateful to anonymous reviewers for valuable~comments.
\section*{Appendix}
In the following table, the coordinates of the discrete charges in Figure~\ref{fig:res} are given for reference.
\vskip 0.2in
\newcolumntype{.}{D{.}{.}{+31}}
\begin{tabular}{| . |.| }
\hline
\multicolumn{1}{|c|}{Electron coordinates} & \multicolumn{1}{c|}{Positron coordinates}\\
\hline
-3.1126741447007592777837555796944 & 2.9530344365605025711769473845442\\
 -2.8775683647214638678316386267493 & -3.0951381769952591512118001250174\\
  -2.6434375832598523581831400447478 & -2.8614098696120820985572576736353\\ -1.9336327385447378036758712466442 & -2.1882892356297157247876068745650\\ -1.6936070421273554090248068336325 & -2.6326516523345556425794778026752\\ -2.1719682464297679903185420586677 & -1.9710273543645197236521756279888\\ -1.4519956803651471460404417607211 & -1.7555235230550466106754786539860\\ -1.2088029171600329445514113518049 & -1.5403236844404032536985003739319\\ -.96395311532928105718506480243987 & -1.1039117628153018022432733933447\\ -.71762228955760307445232301075533 & -1.3237515410575950384914085327471\\ -.47125651467198634491476595790667 & -.87873421856160715534322966231588\\ -.22858211104030609796912348102868 & -.64632138885302414208246702251485\\ .86186952364267439824952086701953 & .33294945631962504080286722164668\\
   .00616783249506095738298247885632 & -.40611981918357315216289810648891\\ .23137130395092737580909635004611 & -.16042959567225122871033380542859\\ .44789490782808160879397312969880& .08676186721721802846687332761214\\ .65746934747098210872577876522952& .57756516301030844804631542526482\\
   1.2615668189213491342215825540197& .82076073717919350910333838443007\\
    1.4597980397241581466451663147530& 1.3035691589932125510920278727478\\ 2.0643955075009215796102922944881& 2.0187559332910846063814218989366\\ 1.6587989484672624969753714473643& 1.7817077278249609050042692119912\\ 1.8599046587703446165936780433590& 1.5432590795101890022517254564541\\ 2.2734402898434140647729061709433& 2.2542171585208840682836715659146\\ 2.7088031793185310430667064345281& 2.7204794077160776393522523646147\\
     2.9363544347316877716974068736781 &\\
\hline
\end{tabular}


\end{document}